\documentclass[english]{./sageo}

\confShortName{SAGEO'2017}
\confLongName{SAGEO'2017 - Rouen, 6-9 novembre 2017}

\journalInEnglishOnly

\usepackage[utf8]{inputenc} 
\usepackage[T1]{fontenc}
\usepackage{lmodern}
\usepackage{textcomp}
\usepackage{amsmath}
\usepackage{graphicx}
\usepackage{multirow}
\usepackage[noend]{algorithmic}
\usepackage[linesnumbered,ruled,vlined,boxed,commentsnumbered]{algorithm2e}

\usepackage{ragged2e}

\firstpagenumber{1}

\title[Spatio-temporal Causalities]{Identification of causalities in spatio-temporal data}

\author[1,2]{Juste}{Raimbault}

\address{UMR CNRS 8504 Géographie-cités}
        {}
\address{UMR-T 9403 IFSTTAR LVMT}
        {juste.raimbault@polytechnique.edu}


\abstract{This paper contributes to the understanding of strongly coupled spatio-temporal processes by describing a generic method based on Granger causality. The method is validated by the robust identification of causality regimes and of their phase diagram for an urban morphogenesis model that couples network growth with density. The application to the real case study of Greater Paris transportation projects shows a link between territorial dynamics, more particularly of real estate and socio-economic, and the anticipated network growth. We finally discuss potential extensions to other temporal and spatial scales.}

\begin{document}

\maketitle


\section{Introduction}

The study of strongly coupled spatio-temporal processes implies to understand tangled intrications generally highly difficult to isolate. These interactions are the essence of complexity approaches, and are indeed at the origin of the emergent behavior of the system. They make sense as an object of study in itself and a separation of processes appears then contradictory with an integrated view of the system. In the case of territorial systems, the example of interactions between transportation networks and territories is a perfect allegory of this phenomenon: methods developed in the seventies aimed at isolating the ``structuring effects'' of a transportation infrastructure~\cite{bonnafous1974methodologies} have later been unveiled as a political instrument and with a poor empirical support~\cite{offner1993effets}. The issue is still highly relevant today as it raises for example with the construction of new High Speed Rail lines in France~\cite{crozethalshs01094554}. The reality of territorial processus is in fact much more complicated than a simple causal relation between the introduction of a new infrastructure and spillovers on local development, but corresponds indeed to complex \emph{co-evolutive} processes~\cite{bretagnolletel00459720}. On long time scales and large spatial scales, some effects of dynamics reinforcement in system of cities by the insertion within networks have been shown by the application of the Evolutive Urban Theory~\cite{espacegeo2014effets}, showing that the disentangling is sometimes possible through a more global understanding of the system. At an other scale, still for relations between networks and territories, we can point at the relations between mobility practices, urban sprawl et ressource localisation in a metropolitan framework~\cite{cerqueira2017inegalites} that are as much complex. This kind of issue is naturally present in other fields: in Economic Geography, the example of links between innovation, local spillovers of knowledge and aggregation of economic agents is a typical illustration of spatio-temporal economic processes exhibiting circular causalities difficult to disentangle~\cite{audretsch1996r}. Specific methods are introduced, as the use of statistical instruments: \cite{aghion2015innovation} shows that the geographical origin of US Congress members that attribute local subsidies is a powerful instrumental variable to link innovation and income inequalities for higher incomes, what confirms that the significant correlation between the two is indeed a causality of innovation on inequalities.

Strong coupling in space and time generally implies a notion of causality, that geography has always studied: \cite{loi1985etude} shows that fundamental issues tackled by contemporary theoretical geography (isolation of objects, link between space and causal structures, etc.) were already implicit in Vidal's classical geography. Beside, \cite{claval1985causalite} criticizes the new determinisms having emerged, in particular the one advocated by some scholars of systemic analysis: in its beginning, this approach inherited from cybernetics and thus of a reductionist vision implying a determinism even for a probabilistic formulation. Claval observes that works contemporary to his writings could allow to capture the complexity that characterizes human decisions: the Prigogine School and the Theory of Catastrophes by René Thom. This viewpoint is extremely visionary, since as Pumain recalls in~\cite{pumain2003approche}, the shift from system analysis to self-organisation and complexity has been long and progressive, and these works have played a fundamental role for it. François Durand-Dastès sums up this picture more recently in~\cite{durand2003geographes}, by focusing on the importance of bifurcations and path-dependency in the initial moments of the constitution of a system that he defines as \emph{systemogenesis}. This type of complex dynamics generally implies a co-evolution of system components, that can be understood as circular causalities between processes: the issue of identifying them is thus crucial regarding the notion of causality for contemporary complex geography.

The regimes under which identification of causalities are relevant are not obviously known. These will depend of the definitions used, as well as available methods for which we give now a few examples. \cite{liu2011discovering} proposes to detect spatio-temporal relations between perturbations of trafic flows, introducing a particular definition of causality based on correspondance of extreme points. Associated algorithms are however specific and difficult to apply to other kind of systems. The use of spatio-temporal correlations has been shown to have in some cases a strong predictive power for trafic flows~\cite{min2011real}. Also in the field of transportation and land-use, \cite{xie2009streetcars} applies a Granger causality analysis, that can be interpreted as lagged correlation, to show for a case study that network growth inducts urban development and is itself driven by externalities such as mobility habits. Neuroscience has developed numerous methods answering similar issues. \cite{luo2013spatio} defines a generalized Granger causality that takes into account non-stationarity and applies to abstracts regions produced by functional imaging. This kind of method is also developed in Computer Vision, as illustrated by \cite{ke2007spatio} that exploits spatio-temporal correlations of forms and flows between successive images to classify and recognize actions. Applications can be quite concrete such as compression of video files by extrapolation of motion vectors~\cite{chalidabhongse1997fast}. In all these cases, the study of spatio-temporal correlations meets the weak notions of causality described above.

In the particular case of relations between network and territories, studies mainly in econometrics have tried to establish causality relationships between variables linked to these two objects. For example, \cite{levinson2008density} explains for the case of London population and connectivity to network variables by these same variables lagged in time, unveiling circular causal effects. \cite{doi10.1068/b39089} uses similar techniques for a region in Italy with historical data on long time, but stays moderate on possible conclusions of systematic effects by recalling the importance of historical events on the estimated relations. \cite{cuthbert2005empirical} proceeds to econometric estimations of reciprocal influence, and concludes that in their Canadian case study at a sub-regional scale, the development of the network induces the development of land-use but not the contrary. Space and time scales influence thus significantly the results of such analysis. \cite{koninghal-00962384} proposes an estimation of relations between the existence of a High Speed Rail connection and economic variables on French Urban Units, and shows a negative effect of the connection itself, after controlling on the endogenous nature of the connection by a selection model, and a significant effect of the characteristics of Urban Units. This work stays limited as it takes neither a time lag larger than one time step nor spatial relations between entities. Finally, still in the same spirit but without explicit inclusion of space, \cite{MANCMANC1073} shows on long time a causality link between infrastructure stock and economic growth on a global panel, but that these effects are moderated locally by under or over-investments.

This contribution aims to explore the possibility of a similar methods for spatio-temporal data exhibiting a priori complex circular causalities, and thus to realize the difficult exercise to couple a certain level of simplicity with a grasping of complexity. We introduce therefore a method to analyse spatio-temporal correlations, similar to a Granger causality estimated in space and time. The robustness of the method is demonstrated in a systematic way by the application to a complex model of simulation of urban morphogenesis, what leads to the unveiling of distinct causality regimes in the phase space of the model. We also include the application to an empirical case study, what positions this work at the interface between knowledge domains of methodology, modeling and empirical within the epistemological framework introduced by~\cite{2017arXiv170609244R}.

The rest of this paper is organized as follows: the generic framework of the method is described in the next section. We then apply it to a synthetic dataset to partially validate it and test its potentialities, what allows us to apply it then to the real case study of Grand Paris transportation network. We finally discuss to proximity with existing methods and possible developments.

\section{Method}

We formalize here the method in a generic way, based in a weak formulation of Granger causality, to try to identify causal relations in spatial systems. Let $X_j(\vec{x},t)$ spatio-temporal unidimensional random processes, which realizations occur in space and time. We give a set of fundamental spatial units  $(u_i)$ that can be for example raster cells or any paving of the geographical space. We assume the existence of functions $\Phi_{i,j}$ allowing to make the correspondance between the realization of each components and spatial units, possibly through a first spatial aggregation or by a more elaborated process driven by a network for example. A realization of a system is given by a set of trajectories for each process $x_{i,j,t}$, and we write a set of realizations $x^{(k)}_{i,j,t}$ (accessible by stochastic repetitions in the case of a model of simulation for example, or by assumption of comparability of territorial sub-systems in real cases). We assume to have a correlation estimator $\hat{\rho}$ applying in time, space and repetitions, i.e. $\hat{\rho}\left[X,Y\right] = \hat{\mathbb{E}}_{i,t,k}\left[XY\right] - \hat{\mathbb{E}}_{i,t,k}\left[X\right]\hat{\mathbb{E}}_{i,t,k}\left[Y\right]$. It is important to note here the hypothesis of spatial and temporal stationarity, that can however easily be relaxed in the case of local stationarity. Furthermore, spatial auto-correlation is not explicitly included, but is taken into account either by the initial spatial aggregation is the characteristic scale of units is larger than the one of neighborhood effects, either by an adequate spatial estimator (weighted spatial statistics of type \emph{GWR}~\cite{brunsdon1998geographically} for example). It allows us to define the lagged correlation by

\begin{equation}
\rho_{\tau}\left[X_{j_1},X_{j_2}\right] = \hat{\rho}\left[x^{(k)}_{i,j_1,t - \tau},x^{(k)}_{i,j_2,t}\right]
\end{equation}

The lagged correlation is not symmetric, but we have directly $\rho_{\tau}\left[X_{j_1},X_{j_2}\right] = \rho_{-\tau}\left[X_{j_2},X_{j_1}\right]$. This measure is applied in a simple way: if $\textrm{argmax}_{\tau} \rho_{\tau}\left[X_{j_1},X_{j_2}\right]$ or $\textrm{argmin}_{\tau} \rho_{\tau}\left[X_{j_1},X_{j_2}\right]$ are ``clearly defined'' (both could be simultaneously), their sign will give the sense of causality between components $j_1$ and $j_2$ and their absolute value the propagation lag. The criteria for significance will depend on the case of application and of the estimator used, but can for example include the significance of the statistical test (Fisher test in the case of a Pearson estimator), the position of extremities of a confidence interval of a given level, or even an exogenous threshold $\theta$ on $\left|\rho_{\tau}\right|$ to ensure a certain level of correlation.

\section{Results}

\subsection{Synthetic Data}

\begin{figure*}[h]
\centering
\includegraphics[width=3.9cm]{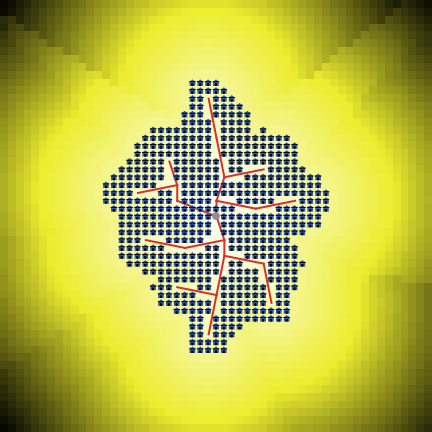}
\includegraphics[width=3.9cm]{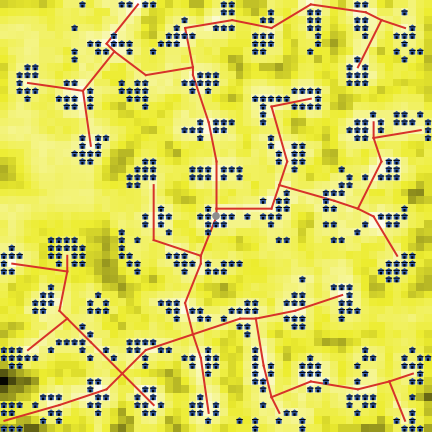}
\includegraphics[width=3.9cm]{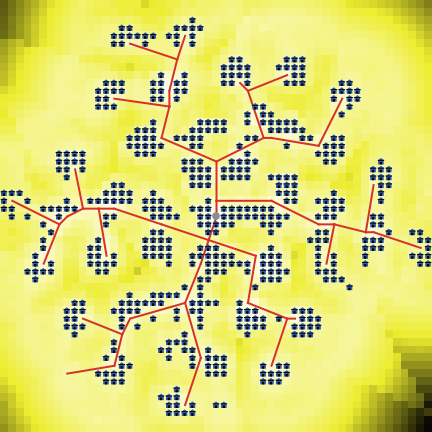}\\\vspace{0.2cm}
\includegraphics[width=12cm]{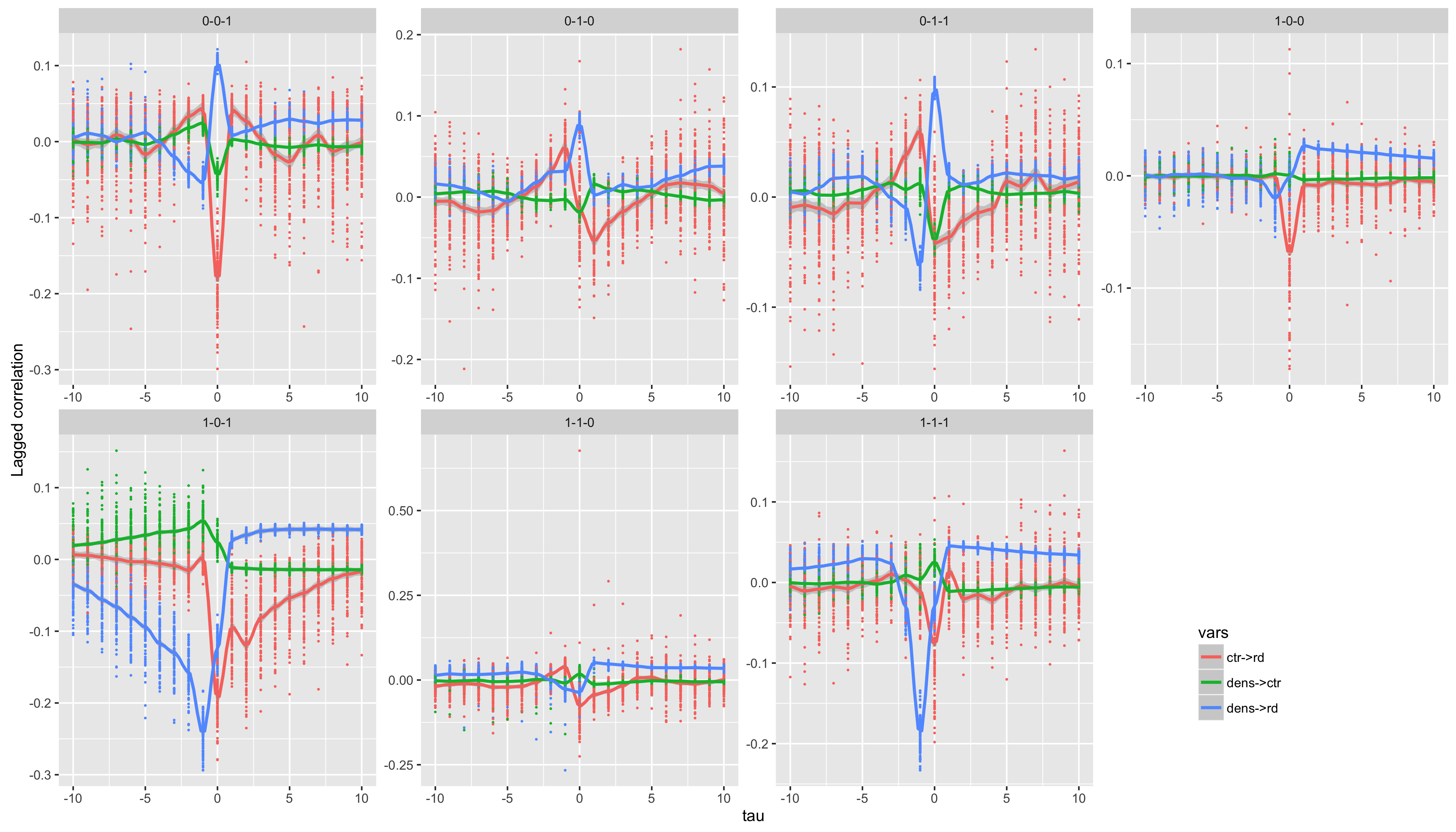}
\caption{\textbf{Correlations in the RBD model.} \textbf{(First row)} Example of different final configurations, obtained with $(w_{d},w_{c},w_{r})$ being respectively $(0,1,1)$,$(1,0,1)$, and $(1,1,1)$. \textbf{(Second row)} Lagged correlations, for each combination of parameters in $\{0,1\}$, as a function of the lag $\tau$. The different colors correspond to each couple of variables: distance to the center (\texttt{ctr}), density (\texttt{dens}) and distance to the network (\texttt{rd}). The dots show the extent on all the repetitions of the model (estimators on $i$ and $t$ only).}
\label{fig:exrdb}
\end{figure*}

\begin{figure*}[h]
\centering
\includegraphics[width=3.9cm,height=3.2cm]{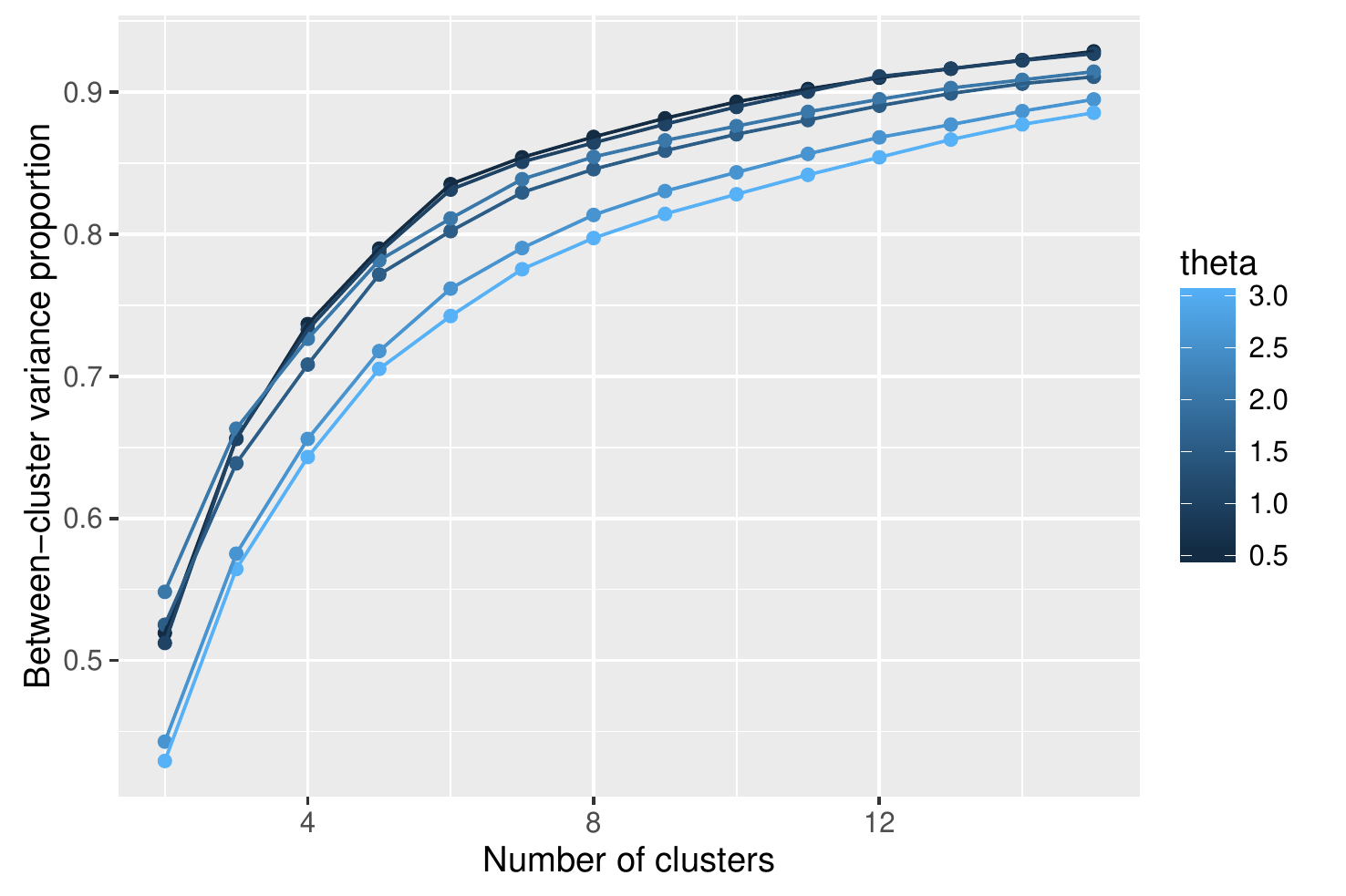}
\includegraphics[width=3.9cm,height=3.2cm]{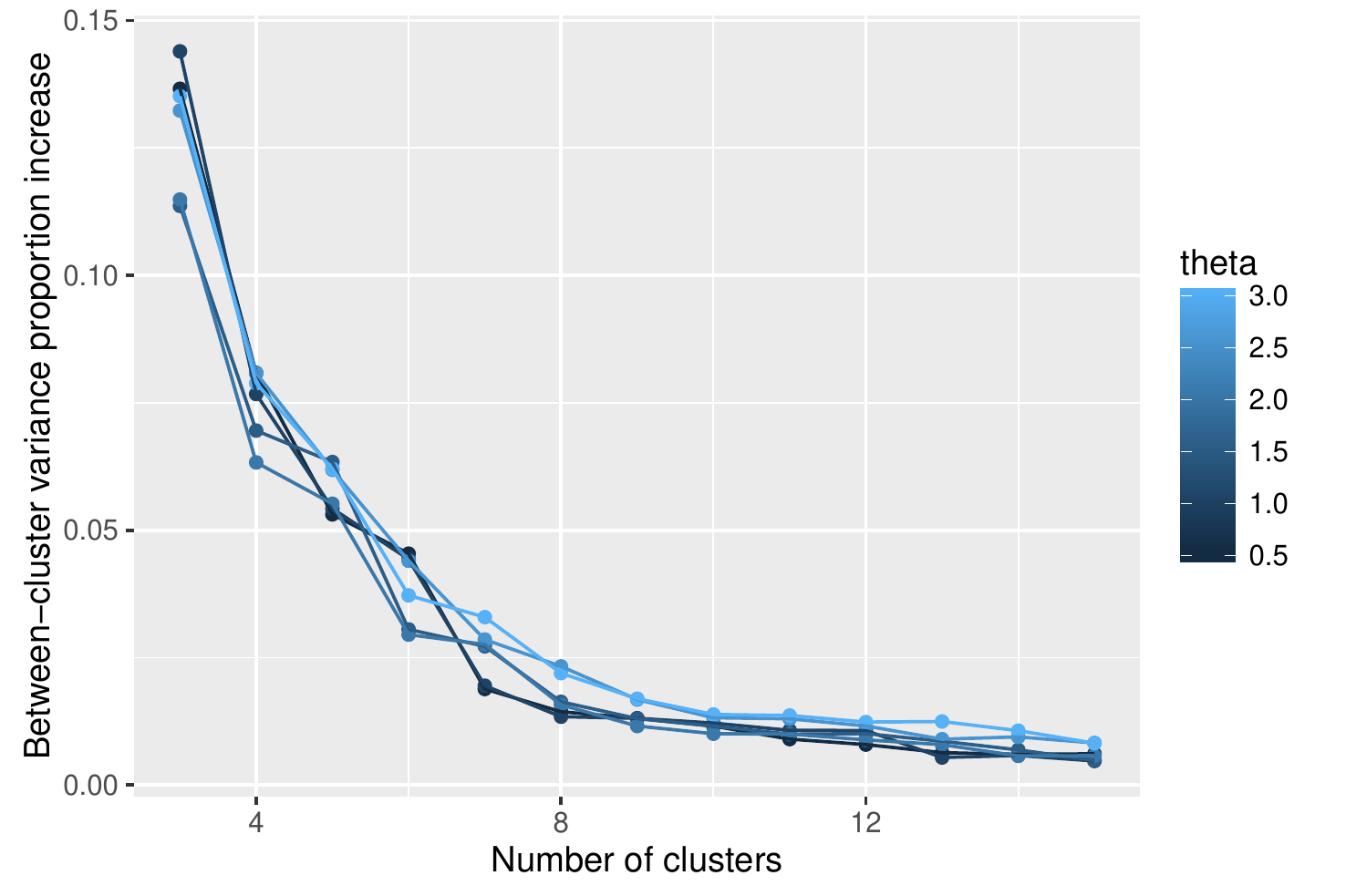}
\includegraphics[width=3.9cm,height=3.2cm]{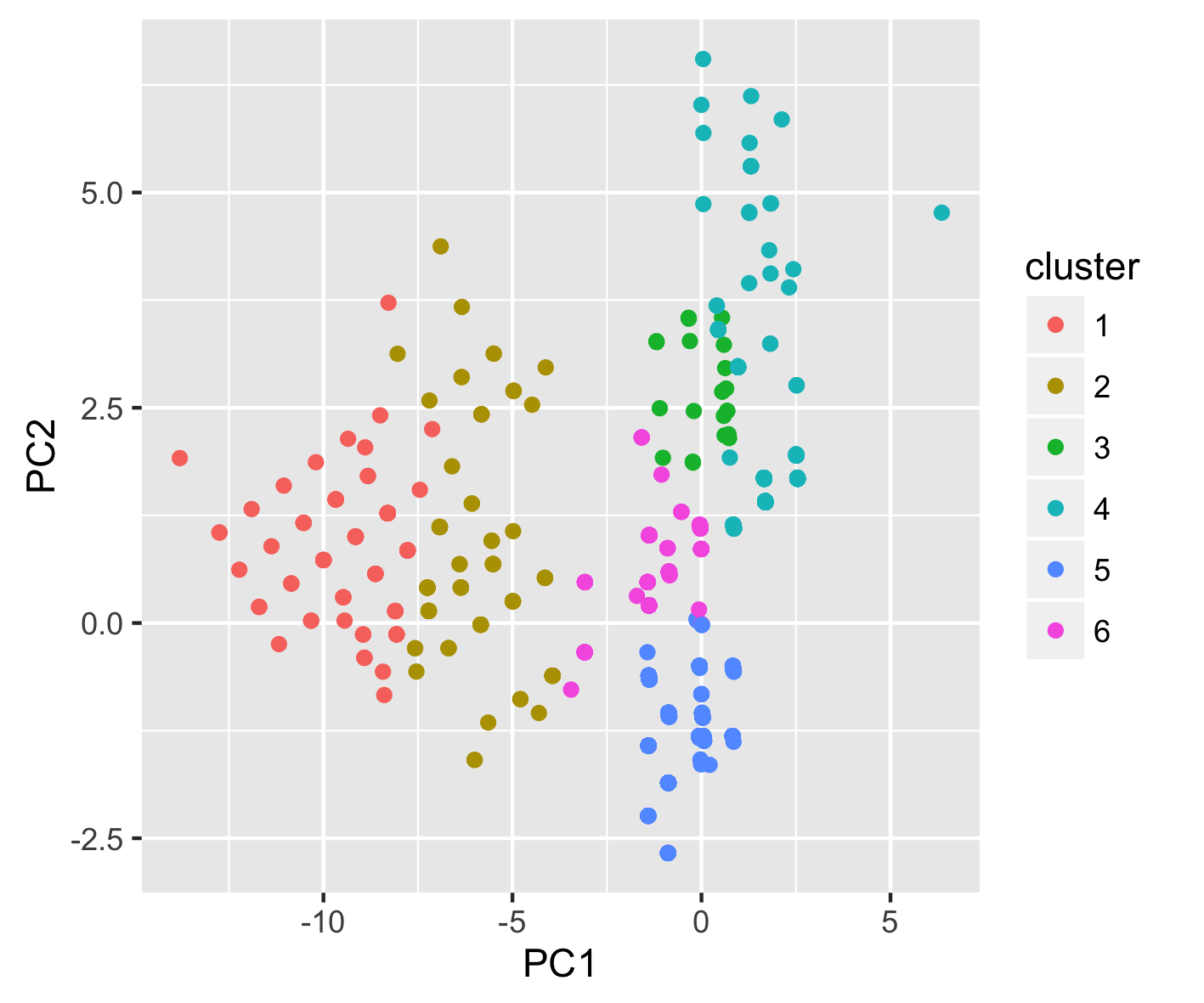}\\
\includegraphics[width=5.9cm,height=5cm]{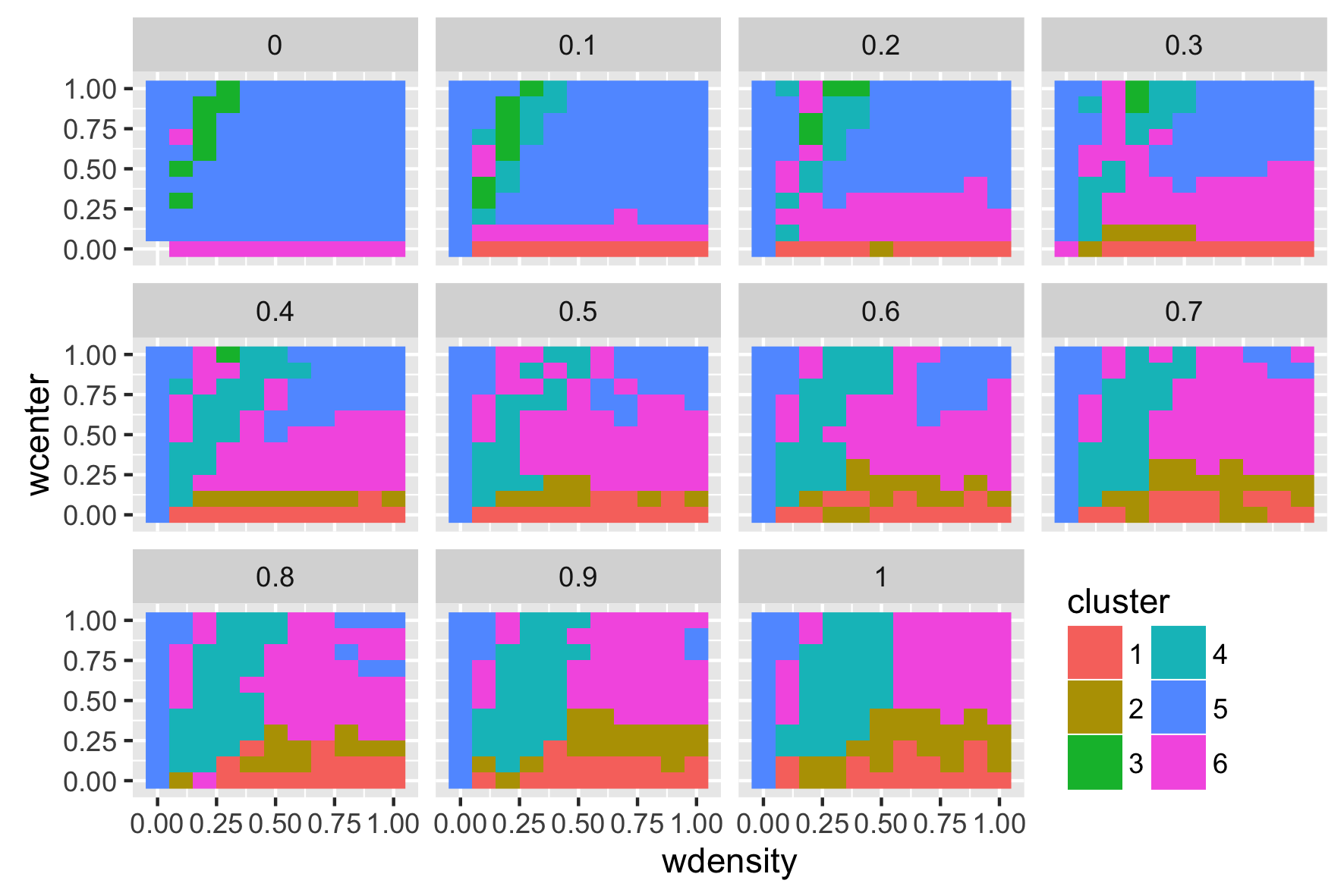}
\includegraphics[width=5.9cm,height=5cm]{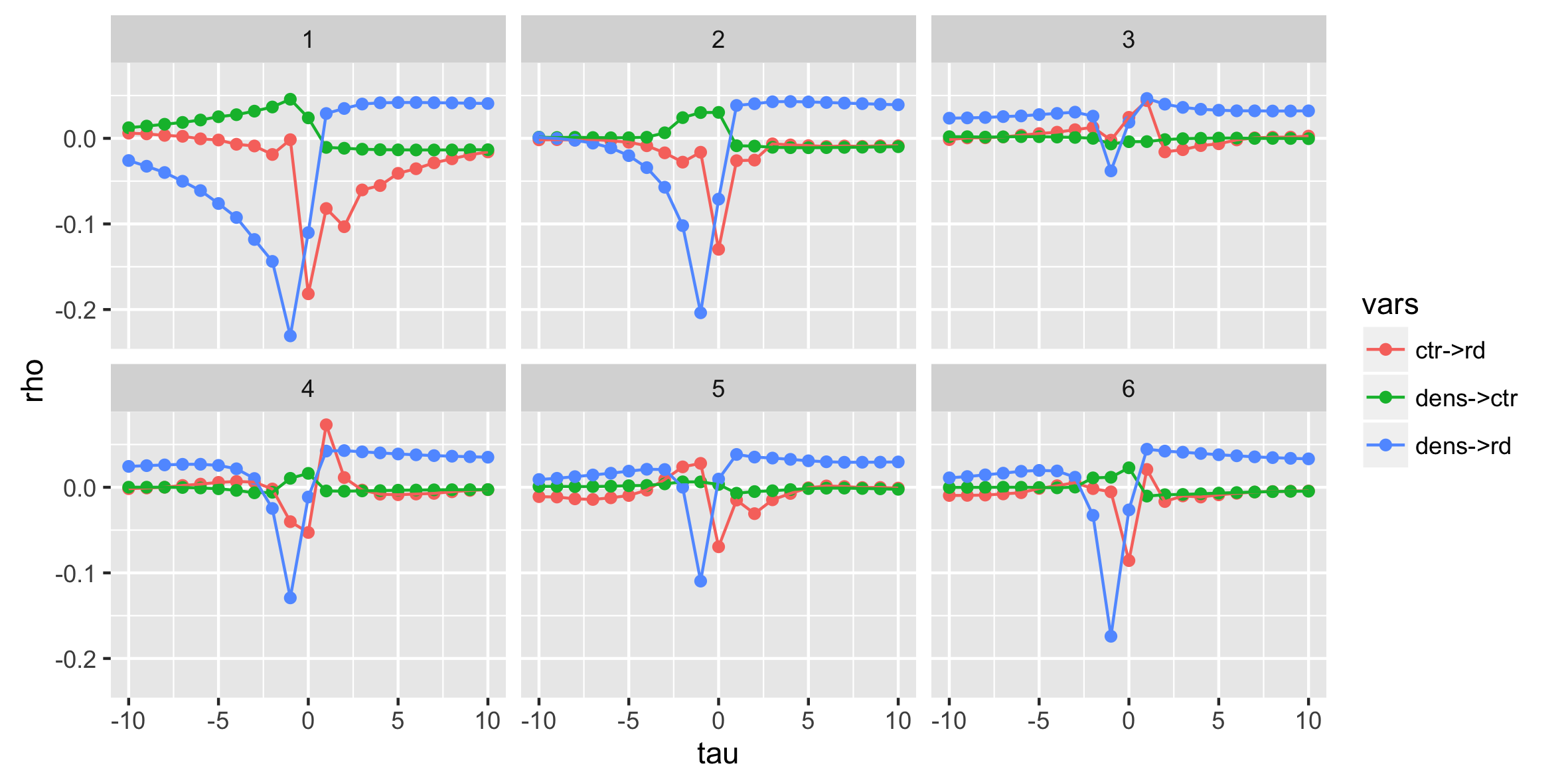}
\caption{\textbf{Identification of regimes of interaction.} \textbf{(Top left)} Inter-cluster variance as a function of cluster number. \textbf{(Top middle)} Derivative of the inter-cluster variance. \textbf{(Top right)} Features in a principal plan (81\% of variance explained by the two first components)\textbf{(Bottom left)} Phase diagram of regimes in the space $(w_{d},w_{c},w_{r})$, $w_r$ varying between the different sub-diagrams of $(w_{d},w_{c}$. \textbf{(Bottom right)} Corresponding profiles of centroids.}
\label{fig:clustering}
\end{figure*}

This method must first be tested and partially validated, what we propose to do on synthetic data, what allows a more refined knowledge of the behavior of models~\cite{raimbaulthalshs01514415}. Echoing the example of relations between transportation networks and territories that introduced the research question before, we generate stylized urban configurations in which network and density mutually interact, and for which causalities are not obvious \emph{a priori} knowing the parameters of the generative model. \cite{raimbault2014hybrid} describes and explores a simple model of urban morphogenesis (the RBD model) that fits perfectly these constraints. Indeed, explicative variables of urban growth, processes of network extension and the coupling between urban density and the network are relatively simple. However, except for extreme cases (for example when distance to the center solely determines land value, the network will depend on density in a causal way; when only the distance to the network counts, the causality will be inverted), mixed regimes do not exhibit obvious causalities. It is for this reason an ideal case to test if the method is able to detect some. We use an applied implementation\footnote{available on the open repository of the project at \\\texttt{https://github.com/JusteRaimbault/CityNetwork/tree/master/Models/Simple/ModelCA}} of the original model, allowing to capture the values of studied variables for each cell of the cellular automaton and for each time step, and to calculate the lagged correlations in the sense described before, between variables of the model. We explore a grid of the parameter space of the RBD model, making the weight parameters for density, distance to center and distance toi network vary\footnote{The model works the following way: a value of cells is determined by the weighted average of these different explicative variables, value that determines the growth of new patches at the next time step.}, that we write respectively $(w_{d},w_{c},w_{r})$, in $\left[0;1\right]$ with a step of $0.1$. Other parameters are fixed to their default values given by~\cite{raimbault2014hybrid}. For each parameter value, we proceed to $N=100$ repetitions, what is enough for a good convergence of indicators. Explorations are done with the OpenMole software~\cite{reuillon2013openmole}, the large number of simulations (1,330,000) implying the use of a computation grid. We compute for all patches the lagged correlations with the unbiased Pearson estimator between the variations of the following variables\footnote{Computing the correlations directly on the variables makes no sense since their value has no absolute meaning.}: local density, distance to center and distance to network. The figure~\ref{fig:exrdb}  shows the behavior of $\rho_{\tau}$ for each couple of variable (undirected, $\tau$ taking negative and positive values), for the combination of extreme values of parameters. We can already see different regimes emerge: for example, $(1,0,1)$ leads to a causality of density on distance to center with a lag $\tau=1$, and a negative causality of density on distance to network with the same lag, whereas distance to the center and to the network are correlated in a synchronous manner. To study these behaviors in a systematic way, we propose to identify regimes endogenously, by using non-supervised classification. We apply a \emph{k-means} clustering, robust to stochasticity (5000 repetitions), with the following features: for each couple of variables, $\textrm{argmax}_{\tau} \rho_{\tau}$ and $\textrm{argmin}_{\tau} \rho_{\tau}$ if the corresponding value is such that $\frac{\rho_{\tau}-\bar{\rho}_{\tau}}{\left|\bar{\rho}_{\tau}\right|} > \theta$ with $\theta$ threshold parameter, 0 otherwise. The inclusion of supplementary features of values of $\rho_{\tau}$ does not significantly changes the results, these are therefore not taken into account to reduce the dimension. The choice of the number of clusters $k$ is generally a difficult problem in this kind of approach~\cite{hamerly2003learning}. In our case the system exhibit an convenient structure: the curves of inter-cluster variance proportion and its derivative in figure~\ref{fig:clustering}, as a function of $k$ for different values of $\theta$, show a transition for $\theta = 2$, what gives for the corresponding curve a break around $k=6$. A visual screening of clusters in a principal plan confirms the good quality of the classification for these values. A class corresponds then to a \emph{causality regime}, for which we can represent the phase diagram as a function of model parameters, and also cluster centers profiles (computed as the barycenter in the full initial space) in figure~\ref{fig:clustering}. The behavior obtained is interesting, as regions in the diagram corresponding to the different regimes are clearly delimited and connected. For example, we observe the emergence of regime 6 in which distance to network causes strongly the density in a negative way, but distance to the center causes distance to the network. Its maximal extent on $(w_d,w_r)$ is for an intermediate value $w_r=0.7$. Thus, to maximize the impact of network on density, the corresponding weight must not be maximized, what can be counter-intuitive at first sight. It illustrates the utility of the method in the case of circular causal relations difficult to entangle a priori. The regime 5, in which distance to network influences the density the same way, but the relation between distance to center and to the network is inverted, is also interesting, and predominates for low $w_r$ values. The regime 1 is an extreme one and corresponds to an isolated situation in which distance to the center has no role: this aspect dominates then totally the other interaction processes between density and network. This application on synthetic data demonstrate on one hand the robustness of the method given the consistence of obtained regimes, and realizes this way a much more finer qualification of model behavior than the one done in the original paper. In this precise case, it can be taken as an instrument of knowledge for relations between networks and territories in itself, allowing the test of assumption or the comparison of processes in the stylized model.

\subsection{Case study}

\subsubsection{Context}

We propose an application on a real case study, still linked to the relations between transportation networks and territories. The metropolitan region of Paris is currently undergoing large mutations, with the institution of a metropolitan governance and new transportation infrastructures for example. The construction of a ring underground allowing suburbs to suburbs links is a rather ancient need, and lead to several proposals on which the State and the Region have been in conflict around 2010~\cite{desjardins2010bataille}. The \emph{Arc Express} project~\cite{stif2007arc} advocated by the Region and focused towards territorial equity can be contrasted with initial proposals for a \emph{Réseau du Grand Paris} aimed at linking ``excellence clusters'' despite a potential tunnel effect. The solution finally chosen (see the last \emph{Schéma Directeur}~\cite{sdrif2013}) is a compromise between the two and allows a rebalancing of accessibility between the west and the east~\cite{beaucire2013grand}. We propose to study the relations between accessibility differentials for each project, with variables linked to real estate and socio-economic variables. Indeed, the links between new lines and real estate value evolution are sometimes dramatic~\cite{damm1980response}.

\begin{figure*}[h]
\centering
\includegraphics[width=12cm]{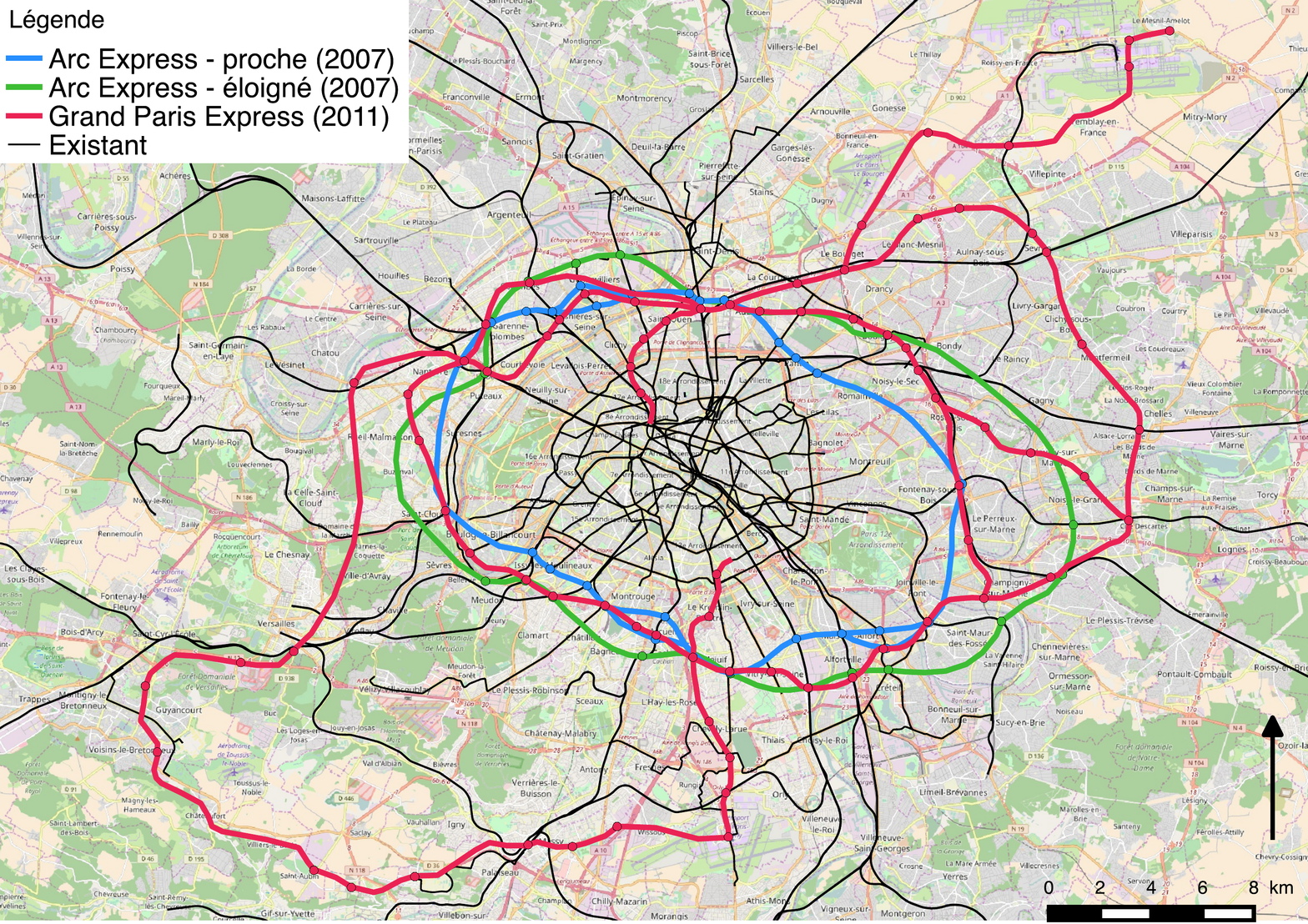}
\caption{\textbf{Successive transportation network projects for the Grand Paris metropolitan area.} We show the two alternatives for the \emph{Arc Express} project elaborated by the Region, and the \emph{Grand Paris Express} (GPE) advocated by the State. The \emph{Réseau du Grand Paris}, a precursor for GPE, is not shown here for visibility reasons because of its proximity with it.}
\label{fig:projects}
\end{figure*}

\subsubsection{Data}

Data for real estate transactions are contained within the BIENS database (\emph{Chambre des Notaires d'Ile de France}, proprietary database). The number of transactions that can be used after cleaning is 862360, distributed across all IRIS areas (basic census units in France), for a temporal span covering the years 2003 to 2012 included. The data at the IRIS level for population and income (median income and Gini index) come from INSEE. Network data have been vectorialized from projects maps (see figure~\ref{fig:projects} for the different projects). Travel times are computed by public transportation only, with standard values for average speeds of different modes (RER 60km.h\textsuperscript{-1}, Transilien 100km.h\textsuperscript{-1}, Metro 30km.h\textsuperscript{-1}, Tramway 20km.h\textsuperscript{-1}). The travel time matrix is computed from all the centroids of IRIS to all the centroids of \emph{Communes} (above aggregation level). These are linked to the network with abstract connectors to the closest station, with a speed of 50km.h\textsuperscript{-1} (travel by car). Analysis are implemented in R~\cite{rcoreteam} and all data, source code and results are available on an open git repository\footnote{At\\\texttt{https://github.com/JusteRaimbault/CityNetwork/tree/master/Models/SpatioTempCausality/GrandParis}. Data for the BIENS database are given only at the aggregated level of IRIS and for price and mortgage variables, for contractual reasons closing the database.}.

\subsubsection{Results}

\begin{figure*}[h]
\hspace{-1cm}
\includegraphics[width=14cm]{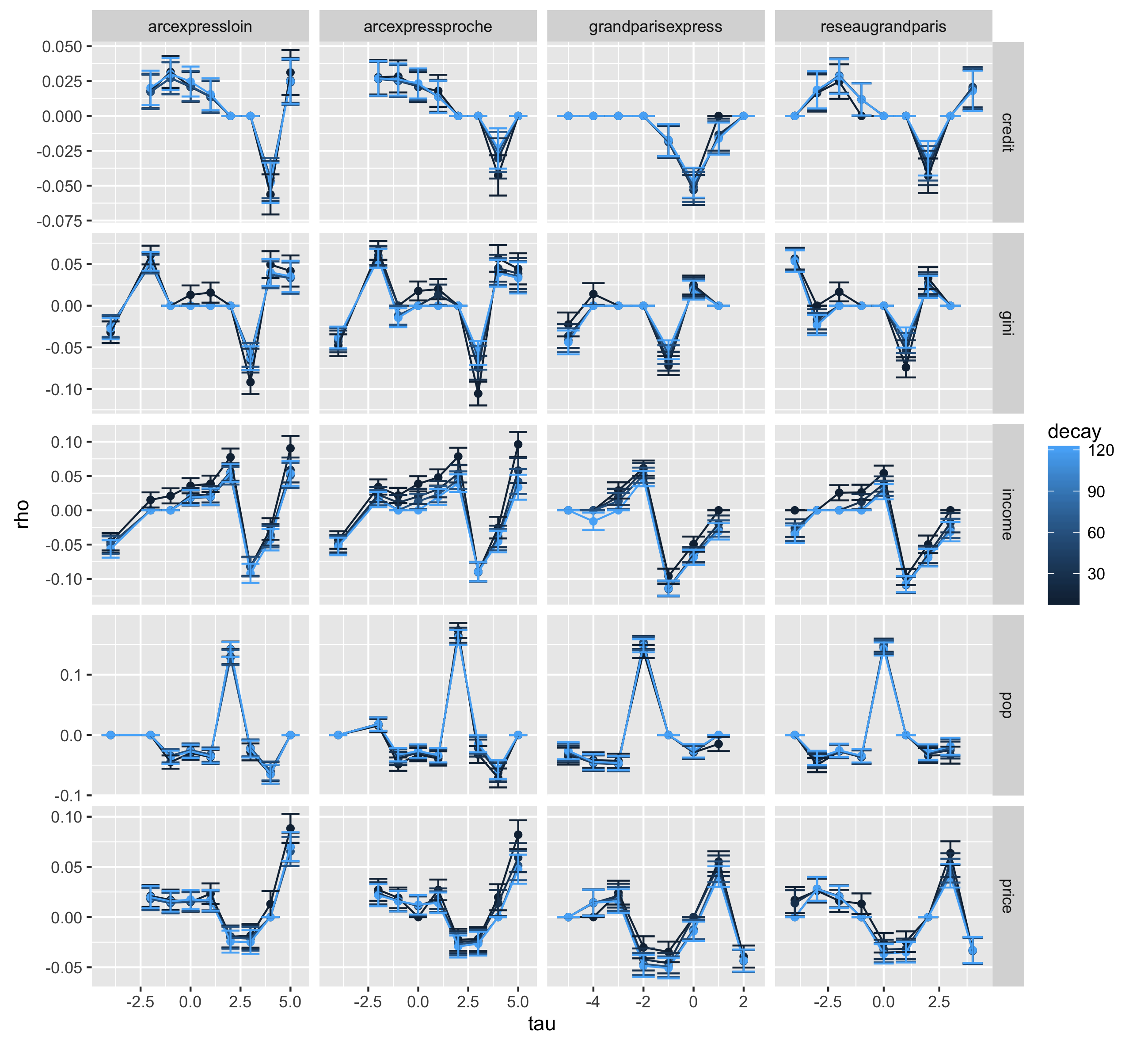}
\caption{\textbf{Empirical lagged correlations.} Plots show the value of lagged correlation between accessibility differentials in average travel time $\Delta T$ for each project (in colunms) and socio-economic and real estate variables variations (in rows). All are computed for different values of the decay parameter (\texttt{decay}, given by curve color). Error bars give the 95\% confidence interval.}
\label{fig:empiricalres}
\end{figure*}

We compute for each project accessibility differentials $\Delta T_i$ in average travel time from each IRIS, in comparison with the network without the project. Average travel time accessibility is defined as $T_i = \sum_k \exp{-t_{ik}/t_0}$ with $k$ \emph{Communes}, $t_{ik}$ travek time, and $t_0$ a decay parameter. To each project is associated a date\footnote{2006 for \emph{Arc Express}, 2008 for \emph{Réseau du Grand Paris} and 2010 for \emph{Grand Paris Express}}, corresponding roughly to the mature announcement of the project. It stays a bit arbitrary as it is difficult on the one hand to determine precisely as a planning project does not emerge from nothing in one day, and one the other hand it may correspond to different realities of learning about the project by economic agents (we do therefore the limiting but necessary assumption of a diffusion of information for the majority of agents in a time smaller than a year). We study the lagged correlations of this variable with the variations $\Delta Y_{ij}$ of the following socio-economic variables: population, median income, Gini index for income, average price of real estate transactions and average value of real estate mortgages. A Fisher test is done for each estimation and the value is set to 0 if it is not significant ($p<0.05$ in a classical manner). The study with generalized accessibility in the sense of Hansen has also been conducted but is less interesting as it has a very low sensitivity to the mobility component (network and decay) compared to the variables themselves. It informs therefore only on relations between these and is not presented here. We show in figure~\ref{fig:empiricalres} the results for all networks and variables. It is first remarkable to note the presence of significant effects for all variables. Lower values for the parameter $t_0$ give correlations higher in absolute value, unveiling a possible higher importance of local accessibility on territorial dynamics. The behavior of population shows a clearly detached peak corresponding to 2008, what suggests an impact of the older project \emph{Arc Express} on population growth, the effect of other projects would then be spurious from their proximity in the most important branches. It would imply that areas where they are fundamentally different such as \emph{Plateau de Saclay} are less sensitive to transportation projects, what would confirm the artificial planned aspect of the development of this territory. Concerning income, we observe a similar behavior but in a negative way, what would imply a decrease of wealth linked to the increase of accessibility, however accompanied by a decrease of inequalities. Finally, real estate prices are as expected driven by the potential arrival of new networks. This effect disappear after two years for the \emph{Grand Paris Express}, suggesting a temporal speculation bubble. We demonstrate thus the existence of complex lagged correlation links, that we call causalities in this sense, between territorial dynamics and anticipated dynamics of networks. A finer understanding of working processes is beyond the scope of this paper and would imply for example qualitative fieldwork or targeted case studies. This example shows however the operational potentialities of our method on a real case study.

\section{Discussion}

\subsection{Spatio-temporal diffusion}

The application of our approach must be lead carefully regarding the choice of scales, processes and objects of study. Typically, it will be not adapted to the quantification of spatio-temporal processes for which the temporal scale of diffusion if of the same order than the estimation window, as our stationarity assumption here stays basic. We could propose to proceed to estimations on moving windows but it would then require the elaboration of a spatial correspondence technique to follow the propagation of phenomena. An example of concrete application that would have a strong thematic impact would be a characterization of a fundamental component of the Evolutive Urban Theory that is the hierarchical diffusion of innovation between cities~\cite{pumain2010theorie}. This would be done by analyzing potential spatio-temporal dynamics of patents classifications such as the one introduced by~\cite{10.1371/journal.pone.0176310}. We also underline that these are rather open methodological questions, for which a concretisation is the potential link between the non-ergodic properties of urban systems~\cite{pumain2012urban} and a wave-based characterization of these processes.

\subsection{Geographically Weighted Regression}

An other direction for developments and potential applications can be found when going to a more local scale, by exploring an hybridation with Geographically Weighted Regression techniques~\cite{brunsdon1998geographically}. The determination by cross-validation of Akaike criterion of an optimal spatial scale for the performance of these models, as done by~\cite{2017arXiv170607467R} in a multi-modeling fashion, could be adapted in our case to determine a local optimal scale on which lagged correlations would be the most significant, what would allow to tackle the question of non-stationarity by a mostly spatial approach.

\section{Conclusion}

We have introduced a generic method of Granger causality on territorial spatio-temporal data, and shown its potentialities and operational nature with synthetic data and on a real case. We postulate that the simple methodological apparel is am asset for a certain level of generality, but that the application to complex case studies exhibiting circular causalities demonstrate the high potential to contribute to the understanding of dynamics for this type of co-evolutive systems.

\section*{Acknowledgements}

Results obtained in the section 3.1 of this paper have been computed on the virtual organisation \textit{vo.complex-system.eu} of the \textit{European Grid Infrastructure} (\texttt{http://www.egi.eu}). We thank the \textit{European Grid Infrastructure} and its \textit{National Grid Initiatives} (\textit{France-Grilles} in particular) to give the technical support and the infrastructure.

\end{document}